\documentstyle[aps,preprint]{revtex}
\begin{document}
\draft
\title{Defect Dynamics for Spiral Chaos in Rayleigh-B\'enard Convection}
\author{M.C. Cross and Yuhai Tu\cite{Present}}
\address{Condensed Matter Physics\\
Caltech, Pasadena CA 91125}
\date{\today }
\maketitle

\begin{abstract}
A theory of the novel spiral chaos state recently observed in
Rayleigh-B\'enard convection is proposed in terms of the importance of {\em %
invasive defects}{\bf \ }i.e defects that through their intrinsic dynamics
expand to take over the system. The motion of the spiral defects is shown to
be dominated by wave vector frustration, rather than a rotational motion
driven by a vertical vorticity field. This leads to a continuum of spiral
frequencies, and a spiral may rotate in either sense depending on the wave
vector of its local environment. Results of extensive numerical work on
equations modelling the convection system provide some confirmation of these
ideas.
\end{abstract}


A novel spatiotemporal chaotic state in Rayleigh-B\'enard convection
consisting of dynamic spirals or targets has recently been discovered
experimentally\cite{M},\cite{S}, and then reproduced numerically first in
equations that approximately model the system\cite{X} and then in the full
thermally driven Navier-Stokes equations\cite{P}. This ``spiral chaos''
state is particularly interesting because it occurs for parameter values
where the familiar state of straight parallel convecting rolls (the
``stripe'' state) is known to be stable, and because the spiral geometry,
although familiar in chemical reaction diffusion systems, was completely
unexpected in this system where the ideal state is static. The spiral state
also provides an intriguing example of the phenomenon of ``defect chaos''
where persistent dynamics is apparently associated with easily identified
defects or coherent structures. Although numerical work has reproduced the
experimental results with remarkable accuracy, there is still little
theoretical understanding why the spiral chaotic state develops, or of the
dynamic behavior of the spiral or target defects within the chaotic state.

In this paper we introduce the idea of an {\em invasive} defect---a
localized structure that through its intrinsic dynamics tends to grow and
take over a significant portion of the system---as a way to understand the
instability (or rather, lack of persistence) of non-ideal states towards the
chaotic state. We analyze the dynamics of individual spirals and targets
based on a slow-variation approach known as ``phase equations'' and show
that ``wave vector frustration'' is crucial to understand this dynamics.
Perhaps surprisingly, fluid vorticity plays a secondary role in the dynamics
of the spirals: the spirals appear to rotate because of the radial motion of
the rolls induced by wave vector frustration exactly as in the radial motion
of the axisymmetric target states. This leads us to a prediction for when
the spiral chaos state should be expected. We analyze the transition between
a spiral dominated state and a target dominated state in terms of a novel
core instability of an axisymmetric target to nonaxisymmetric perturbations.
In addition we present results of extensive numerical investigations of
classes of model equations that allow us to test these ideas, and for the
first time numerically demonstrate the spiral to target transition.

Our theoretical analysis is based on the ``phase equation'' \cite{CN}---a
dynamical equation for a variable describing displacements of a local stripe
pattern. An important feature of the convection system is that this dynamics
is coupled to an additional long wavelength mode consisting of a horizontal
fluid flow with nonzero mean averaged over the depth of the cell. Such a
flow is induced by distortions of the convective rolls (e.g. curvature) and
in turn advects the rolls. Since the mean flow must be divergence free by
the near incompressibility of the fluid, it takes the form of a stirring
motion, and may be completely described in terms of a vertical vorticity $%
\Omega $. Previous work \cite{X},\cite{P} on the spiral chaos has suggested
that the vertical vorticity resulting from the chiral nature of the spiral
state may be important in the observed dynamics of the spirals. We find that
the role of $\Omega $ is more subtle than this.

The phase variable $\theta (\vec x,t)$ is defined as a function of time and
the horizontal coordinate $\vec x=(x,y)$ so that the local wave vector is $%
\vec q=\vec \nabla \theta $. An expansion in slow variations of a basic
stripe pattern yields the equation\cite{CN}
\begin{equation}
\dot \theta +\vec U\cdot \vec q=\tau ^{-1}\left( q\right) \vec \nabla \cdot
\left( \vec qB(q)\right)  \label{phase}
\end{equation}
where $\vec U$ is the divergence free mean flow completely characterized by
the vertical vorticity $\Omega =\hat z\cdot \vec \nabla \times \vec U$ in
turn driven by distortions of the stripes
\begin{equation}
\Omega =-\gamma (q)\hat z\cdot \vec \nabla \times \left[ \vec q\vec \nabla
\cdot \left( \vec qA^2(q)\right) \right]  \label{vorticity}
\end{equation}
In these expressions $\tau ,\gamma$, $A$ and $B$ are functions of the magnitude
of the local wave vector and depend on the control parameter as well as
other fluid parameters. In (\ref{vorticity}) we have used the simplified
form valid near threshold \cite{CN}: the complicated expression valid more
generally\cite{NP},\cite{CH} should not change the conclusions.

{}From these equations we show that the spiral dynamics is induced by the
radial motion of rolls forced by wave vector frustration rather than by a
rotation of the structure driven by a vorticity field induced by the chiral
structure. An important consequence of this is that a spiral of a given
chiral sense may rotate in {\em either} direction. In addition the spiral
frequencies in a system will form a continuum rather than a discrete set or
unique value as found for chemical spirals \cite{CH}.

This result is readily obtained by integrating (\ref{phase}) for the phase
of a spiral $\theta =k(r)r+m\phi -\omega t$ over a cylindrical geometry
radius $R$ with $R>>1$ for a well developed spiral:
\begin{equation}
\omega \int d^2r\;\tau (q)=\int d^2r\;\tau (q)\vec U\cdot \vec q-\left. \int
B(q)q_rrd\phi \right| _{r=R}  \label{frequency}
\end{equation}
(Here $r$ and $\phi $ are cylindrical polar coordinates.) The second term on
the right, which is also present in the axisymmetric target case, is zero if
$q=q_f$, the ``focus selected wave number''\cite{PM} defined by $B(q_f)=0$%
\cite{CN}, and is nonzero only if the wave number is held away from this
value by some other means (wave vector frustration). In this case the
contribution to $\omega $ is of order $\tau ^{-1}B^{\prime
}(q_f)q_f\,(q-q_f)R^{-1}$ with $\tau $ an average of $\tau (q)$ over the
cell. We can estimate the size of the first term by correspondingly
integrating the vorticity equation \ref{vorticity}
\begin{equation}
\int d^2r\;\left[ \gamma (q)\right] ^{-1}\Omega =-\oint_R\vec \nabla \cdot
\left( \vec qA^2\right) \vec q\cdot d\vec l+\oint_{r\rightarrow 0}\vec \nabla
\cdot \left( \vec qA^2\right) \vec q\cdot d\vec l
\end{equation}
It is readily checked that for $R\gg 1$ where $A^2\rightarrow const$. the
first term on the right hand side goes as $R^{-1}$. If $A$ remained constant
as $r\rightarrow 0$, the second term would correspondingly diverge as $%
r^{-1} $. Instead, of course, $A\rightarrow 0$ over the coherence length $%
\xi $ for spatial variations of the amplitude of the pattern. Thus the
integrated vorticity is dominated by a contribution localized in the core
region where amplitude and phase are rapidly varying, rather than a
contribution from the slowly varying spiral arm structure at large
distances. Now $\vec U\cdot \vec q\simeq U_\phi m/r\sim m\bar \Omega \xi
^2/r^2$ with $\bar \Omega $ an average vorticity, so that the first term in (%
\ref{frequency}) gives a contribution to the frequency $\omega $
proportional to $R^{-2}$, smaller by the factor $R^{-1}$.

The invasive nature of spirals and targets now rests on the comparison of
the background wave number $q_b$ in their vicinity with $q_f\,$: if $q_b>q_f$
then rolls will tend to move into the defect center, a spiral will
``unwind'' and, unless there is some coherent source of these rolls
elsewhere, the background state will tend to take over the spiral or target;
on the other hand if $q_b<q_f$ the target or spiral will tend to invade the
background. The question of what to assume for $q_b$ is quite hard, and
there may not be a completely general argument. We propose that since the
background state often contains many stationary or slowly moving roll
dislocations, an appropriate choice for $q_b$ is the value $q_d$ of the wave
number at which an inserted dislocation has zero velocity along the rolls
(climb velocity)\cite{PMd},\cite{TC}. This leads to the prediction that
spiral or target chaos is expected for $q_d<q_f$.

The combined motion of a spiral and dislocation can be understood more fully
by considering the situation shown in Fig.\ref{fig1} with a single
dislocation terminating a spiral at radius $R$. The motion can be split into
two components. There is an outward motion of the rolls, driven by the
curvature, at a speed $\alpha (q_f-q)/qR$ with $\alpha $ a positive
proportionality constant and $q$ the wave vector at the dislocation: this
would leave the spiral as shown by the dotted line. On the other hand the
dislocation will climb at a speed $\beta (q-q_d)$ moving in the azimuthal
direction and shortening the spiral, with $\beta $ another positive
proportionality constant. A steady state solution in which the spiral
rotates at constant rate and with constant length exists if these effects
cancel i.e. if $q=(\alpha q_f+\beta q_d)/(\alpha +\beta )$, a weighted mean
of $q_f$ and $q_d$.

We have investigated this picture in numerical simulations of equations
modelling the convection system as in the work of Xi et al.\cite{X}. In this
approach the system is described by an equation for a real field $\psi (\vec
x,t)$, a function of horizontal coordinates and time only, that represents
the horizontal structure of the convection pattern. The dynamical equation
for $\psi $ is based on the Swift-Hohenberg equation\cite{SH} that has been
much used in the study of stationary patterns and transient relaxation
processes\cite{CH}:
\begin{equation}
\dot \psi +\vec U\cdot \vec \nabla \psi =\epsilon \psi +\left( \nabla
^2+1\right) ^2\psi -g\psi ^3+3(1-g)(\nabla \psi )^2\nabla ^2\psi   \label{SH}
\end{equation}
where again $\vec U$ is the vorticity induced mean flow. For $g=1$ (\ref{SH}%
) reduces to the Swift-Hohenberg model. The additional nonlinear term \cite
{GC} for $g\not =1$ yields a more accurate reproduction of the fluid
stability balloon and provides an additional tuning parameter to investigate
transitions in behavior. We have investigated various forms for the driving
of the vorticity by distortions of the field $\psi $: the first (``dynamic
vorticity'') follows Xi et al.\cite{X}
\begin{equation}
\dot \Omega -\sigma (\nabla ^2-c^2)\Omega =g_m\;\hat z\cdot \vec \nabla
(\nabla ^2\psi )\times \vec \nabla \psi   \label{DV}
\end{equation}
where $\sigma $ is the fluid Prandtl number and $g_m$ determines the
strength of the coupling between the mean flow and $\psi $. We have also
considered two modifications to (\ref{DV}). The first (``passive
vorticity'') is to eliminate the term $\dot \Omega $ term, which together
with $\nabla ^2\Omega $ is higher order in the gradient expansion leading to
(\ref{DV}). The second is to reduce the high wave vector Fourier components
of the vorticity by including the ``filtering operator'' $F_\gamma $
introduced in ref.\cite{GC}.This term reduces the short wavelength
components of the vorticity field, suppressing an additional instability in
Eqs.(\ref{SH}),(\ref{DV}) that does not correspond to the behavior of the
fluid system c.f. ref. \cite{P}.

We have fixed $\epsilon =0.7$, $\sigma =1$, $c^2=2,$ as in ref. \cite{X},
and have varied the vorticity coupling $g_m$, the non-linearity coefficient $%
g$, and the form of the vorticity equation. Our simulations are done in a
square geometry with periodic boundary conditions using the same numerical
scheme as in ref. \cite{CTM}. A dynamic spiral or target state rapidly forms
from random initial conditions over a wide range of parameters. We indeed
see spirals rotating in either direction relative to their chirality,
although since the unwinding sense is observed to lead to their destruction,
the winding sense is more prevalent.

We have studied the defect chaos state as a function of $g_m$ and $g$ to
assess the importance of $q_f$ and $q_d$. Fig.\ref{spirals} shows the
spatial structure a time 4000 after random initial conditions in a size $%
200\times 200$ using dynamic, unfiltered vorticity. Panel (a) shows the
dynamic spiral state for $g_m=50$ and $g=1$. As $g_m$ is reduced targets
become more prevalent e.g. panel (b) with $g_m=10$. Finally as $g_m$ is
reduced further, panel (c) with $g_m=2$, the target-spiral state disappears,
and the pattern freezes into a disordered stripe state. If alternatively $g$
is reduced changing the form of the nonlinearity, the spatial structure
remains rather unchanged, until at low $g$ again the spiral structures
disappear, but now to a new dynamic state in which dislocation pairs,
evident in panel (d) with $g=0.2,$ rapidly migrate across the system \cite
{filtered transition}. The quantitative aspects of the evolution of the
structures with $g_m$ and $g$ are shown in Fig.\ref{g-gm}. Both as a
function of $g_m$ and $g$ the mean wave vector is seen to follow a trend
consistent with a weighted mean of $q_f$ and $q_d$, except for the lowest $%
g_m$ where in fact the spatial structure shows few spirals or targets. In
addition the spiral/target state disappears when $q_d$ approaches $q_f$,
although we do not have a model for which the $q_f$ and $q_d$ lines actually
cross. It should also be noted that the wave vector distribution is not
determined by the stability boundaries. This is shown for example in (b)
where the distribution varies with $g_m$ whereas the Eckhaus stability
boundary does not. The irrelevance of the cross-roll instability is shown by
similar results (not shown) for the model with filtered vorticity when the
CR line is suppressed to larger wave vectors.

Further insight into the spiral-target transition is given by considering
the dynamics of a single target in an axisymmetric geometry. Fig.\ref{core}
shows time independent solutions constrained to axisymmetry in a circular
geometry with boundary conditions $\psi =\partial \psi /\partial r=0$ that
act to pin the phase of the rolls at the outer boundary. As the radius is
increased, the rolls are stretched and the value of the field at the center
goes up/down, until eventually a new roll is nucleated at the center. For
the parameters used in Fig.\ref{core} this event is hysteretic: if the
radius is decreased the roll disappears at a smaller radius than the one at
which the nucleation occurred. Numerical linear stability analysis shows
that the portion of the curve with filled symbols in Fig.\ref{core} is
unstable to an $m=1$ symmetry breaking perturbation (the most unstable
mode). The eigenvector of the instability, (inset to Fig.\ref{core}), shows
it to be localized at the core of the target, in contrast to the instability
studied by Newell et al. \cite{Newell}. As $g_m$ is decreased, the
instability point $P$ moves closer to the point where the hysteretic jump
occurs, and reaches this point at $g_m\simeq 21.5$.

In the dynamics of a target in a disordered background (e.g. Fig.\ref
{spirals}) we can suppose that the target will approximately pass through
the sequence of states represented in Fig.\ref{core}a, with roll radiation
corresponding to increasing the system length. We cannot say exactly where
the jump from one branch to the other, corresponding to the nucleation of a
new roll) will occur. However we see that if the nucleation occurs
sufficiently early the core may experience the nonaxisymmetric instability
providing $g_m$ is greater than about $20$. We might guess that this would
then lead to a spiral core structure, which would further evolve to convert
the target to a spiral. In fact our numerics of the disordered state in
periodic boundary conditions shows that the subsequent nonlinear behavior is
more complicated: in the target state at $g_m=20$ for example, we observe a
complicated, almost periodic behavior in which the nonaxisymmetric
perturbation successively grows and decays over many cycles ($\sim 15$)
before a new roll is finally nucleated and the structure regains the form of
a target.

In conclusion, we have proposed that the novel spiral/target chaos state may
be understood in terms of the invasive dynamics of these defects. This
dynamics in which rolls are radiated from the center is driven by wave
vector frustration. For the spirals the rotation induced by the vertical
vorticity plays a secondary role in the dynamics. The invasive dynamics
depends on the balance between the focus selected wave vector $q_f$ and the
background wave vector. The vorticity is however important in generating the
spiral core, and the target to spiral transition may be partially understood
in terms of a nonaxisymmetric instability of the axisymmetric core structure
of a target. In addition, in the absence of the coupling to the vertical
vorticity the focus selected wave vector $q_f$ is marginally unstable to a
zigzag distortion, so that the vorticity may also be important in
stabilizing the far field of the targets and spirals, as well as in the
motion of the dislocations and hence the value of $q_d$. Our proposal
provides a criterion for when spiral chaos may be expected. The weakest part
of our analysis is the identification of the background wave vector with $%
q_d $: alternative possibilities might be a wave vector selected by another
defect in the system that acts as a sink of the rolls, or a wave vector at
which a roll destruction process develops, for example by the production of
dislocation pairs or the nucleation of a mobile dislocation from the core of
a stationary defect.

This work was supported by the NSF through grant DMR-9311444 and by the San
Diego Supercomputer Center.

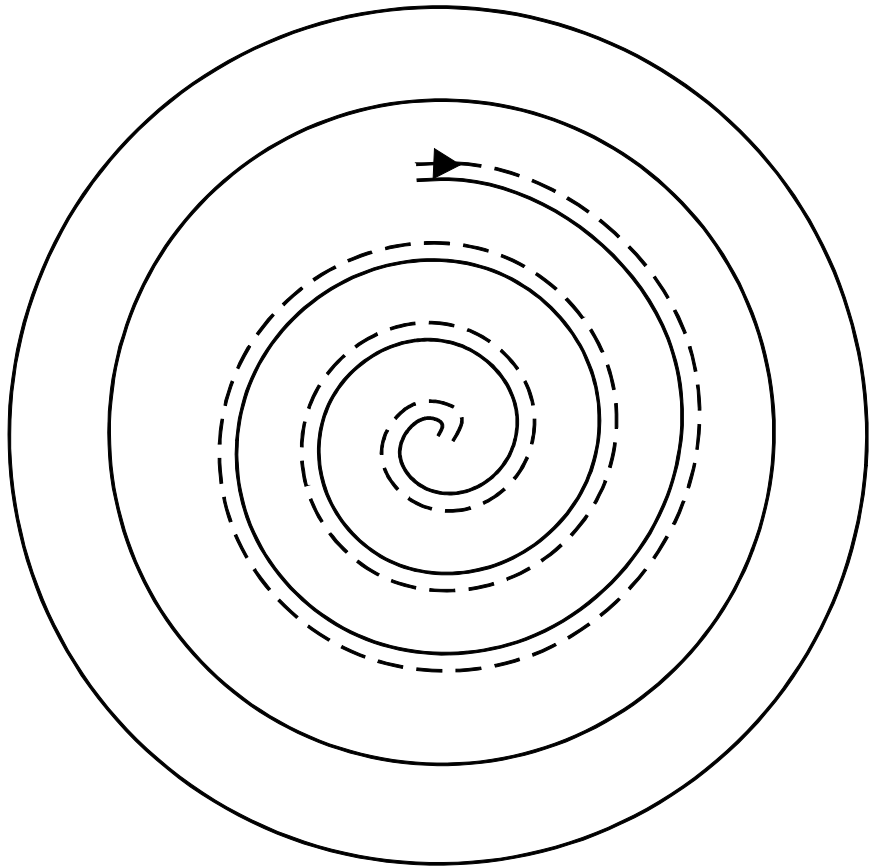
\begin{figure}
\caption{Spiral defect ending in a dislocation}
\label{fig1}
\end{figure}
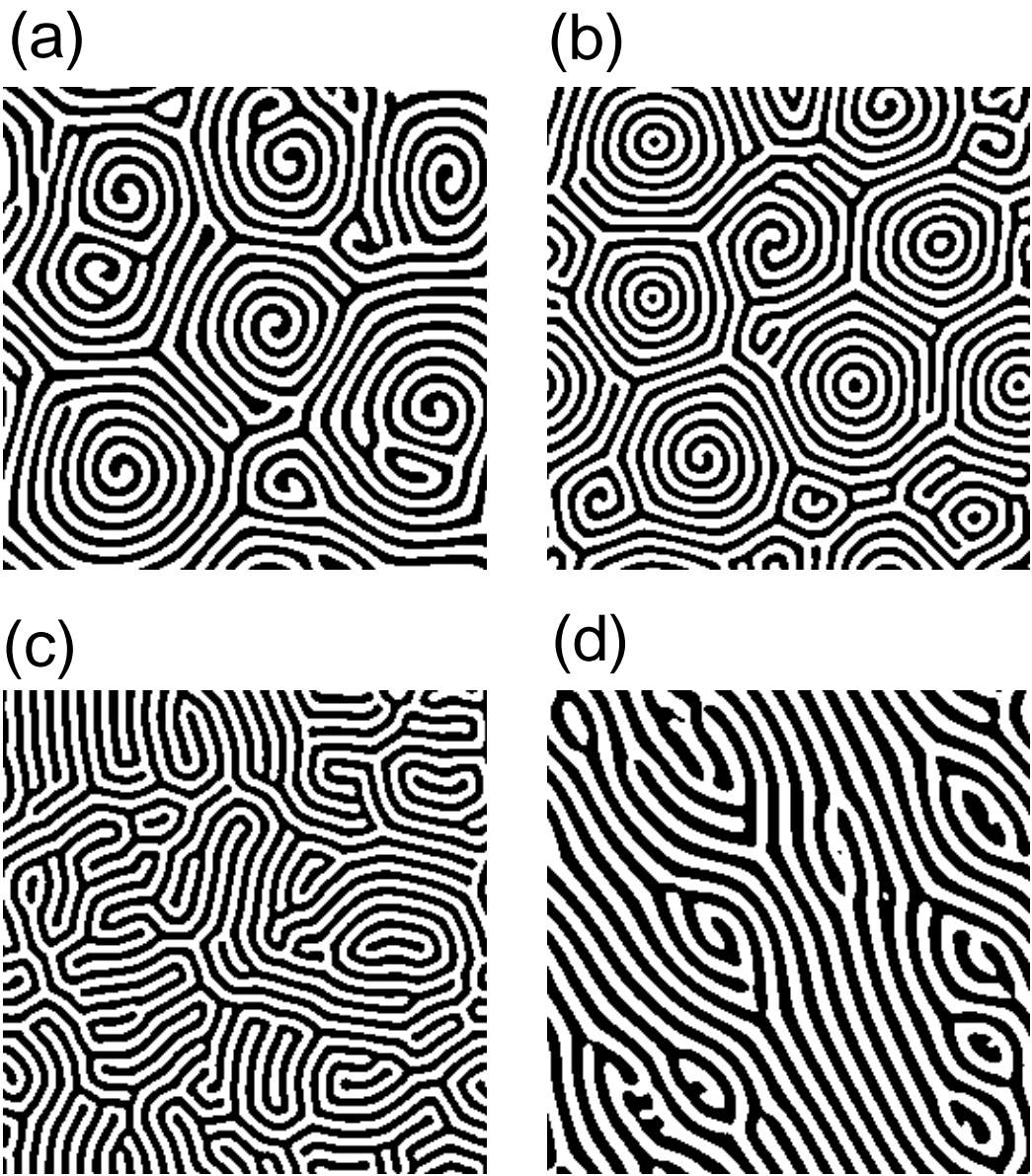
\begin{figure}
\caption{Disordered states a time 4000 after random initial conditions in a
system of size $200\times 200$ for dynamic unfiltered vorticity showing the
variation of the patterns as $g$ and $g_m$ are varied. Values are (a) $g=1$,
$g_m=50$; (b) $g=1$, $g_m=10$; (c) $g=1$, $g_m=2$; (d) $g=0.2$, $g_m=50$.}
\label{spirals}
\end{figure}
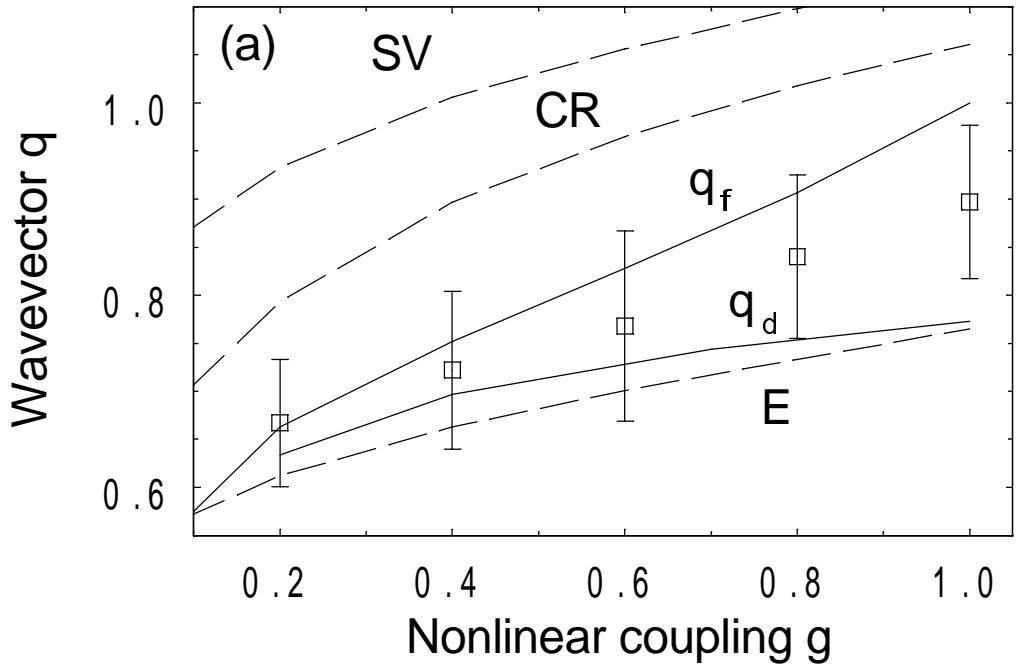
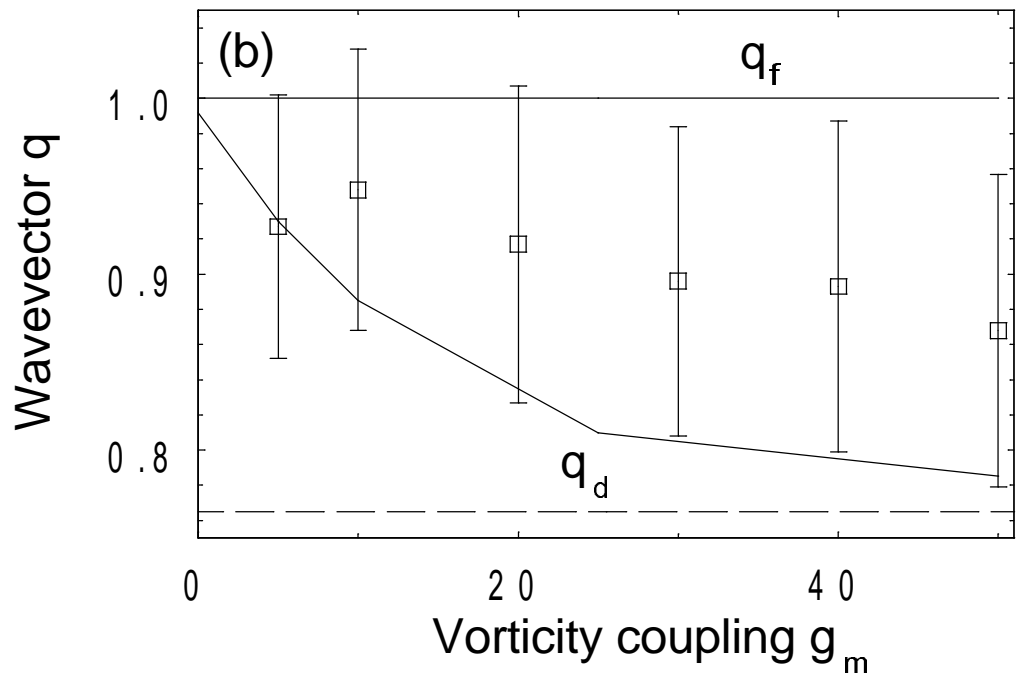
\begin{figure}
\caption{Wavenumber distribution in disordered states as in fig.(\protect\ref
{spirals}) as a function of (a) $g$ and (b) $g_m$. The symbols are the mean
wavevector and the error bars give the standard deviation, both given by a
gaussian fit to the distribution. The lines show $q_f$ and the wavevectors
for the charateristic instabilities of straight parallel rolls (E: Eckhaus;
SV: skew-varicose; CR: cross-roll) \protect\cite{HSG}, and $q_d$ given by
separate numerical calculations of the velocity of climb of a pair of
dislocations in straight roll backgrounds of various wave vectors, as in
ref.\protect\cite{TC}.}
\label{g-gm}
\end{figure}
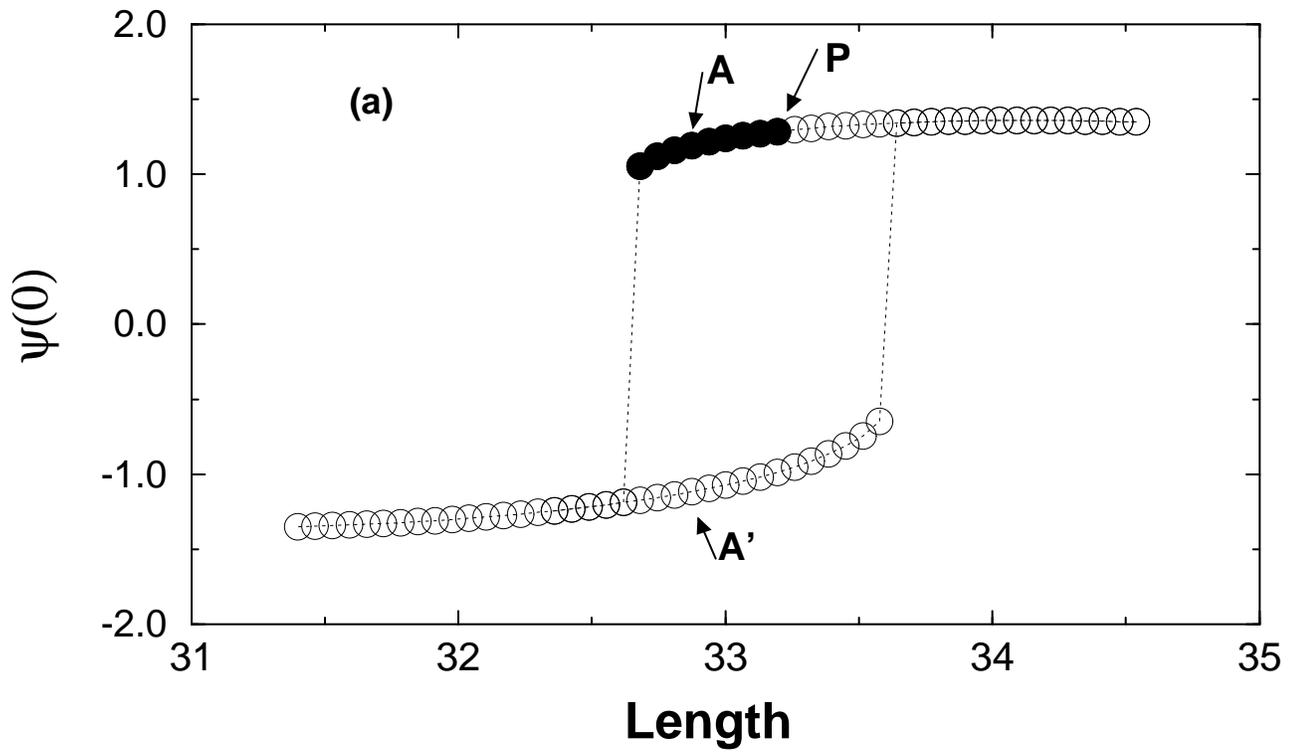
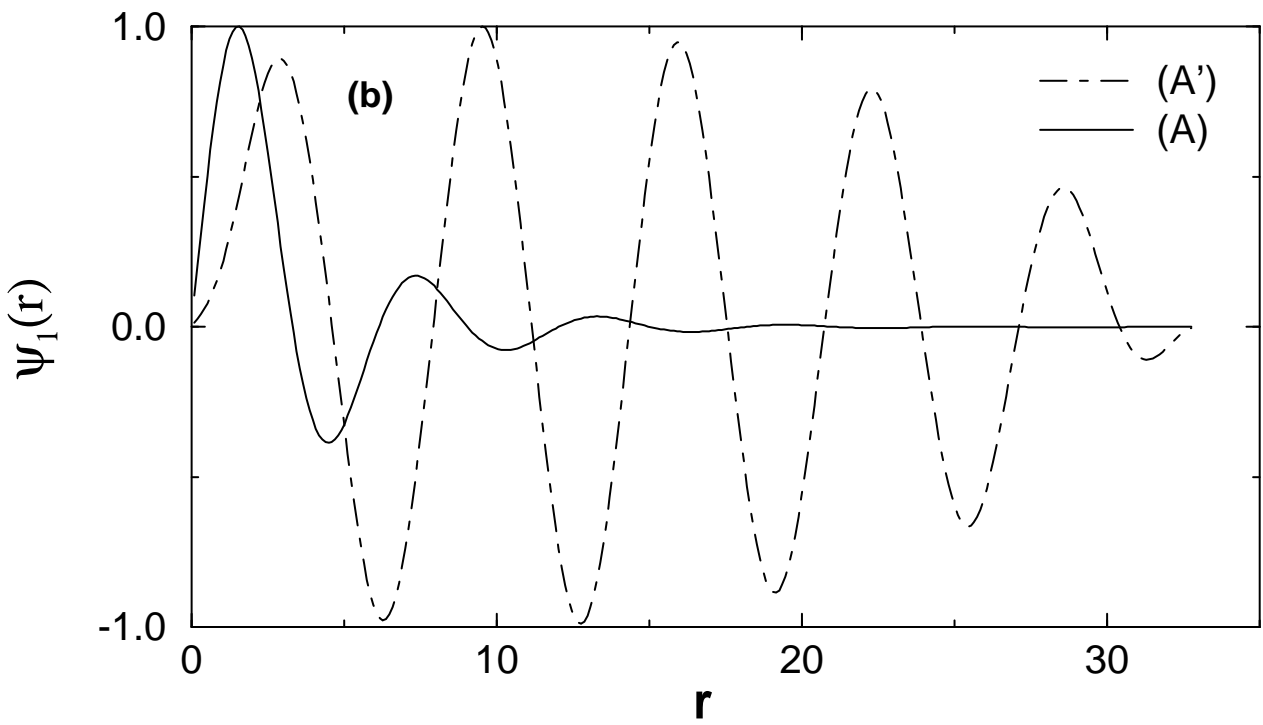
\begin{figure}
\caption{(a) Variation of the field $\psi (r=0)$ at the center of a target as
a function of the radius of the system. Notice the hysteretic transition. The
states represented by the filled symbols are unstable to an $m=1$ distortion.
(b) radial dependence of the eigenvector of the unstable mode at the point $A$
as well as the stable eigenvector with largest eigenvalue at the corresponding
point $A^{\prime }$ on the stable branch. Notice the
localized nature of the instability. Parameters are $g=1$, $g_m=50$,
$\epsilon=0.7$.}
\label{core}
\end{figure}


\begin{references}
\bibitem{filtered transition}  It is interesting that with filtered, passive
vorticity the transition as $g$ is varied appears more continuous, with
increasing swaths of straight rolls between more and more dilute defects.
\end{references}
\end{document}